\documentclass[twocolumn,showpacs,preprintnumbers,amsmath,amssymb]{revtex4}

\usepackage{graphicx}
\usepackage{dcolumn}
\usepackage{bm}

\begin{document}


\title{Experimental Quantum Key Distribution
with Decoy States}

\author{Yi Zhao, Bing Qi, Xiongfeng Ma, Hoi-Kwong Lo, Li Qian}
\affiliation{Center for Quantum Information and Quantum Control,
Department of Physics and Department of Electrical \& Computer
Engineering, University of Toronto, Toronto, Ontario, M5S 3G4,
Canada
}%

\date{\today}

\begin{abstract}
To increase dramatically the distance and the secure key generation
rate of quantum key distribution (QKD), the idea of quantum
decoys---signals of different intensities---has recently been
proposed. Here, we present the first experimental implementation of
decoy state QKD. By making simple modifications to a commercial
quantum key distribution system, we show that a secure key
generation rate of 165bit/s, which is 1/4 of the theoretical limit,
can be obtained over 15km of a Telecom fiber. We also show that
with the same experimental parameters, not even a single bit of
secure key can be extracted with a non-decoy-state protocol.
Compared to building single photon sources, decoy state QKD is a
much simpler method for increasing the distance and key generation
rate of unconditionally secure QKD.
\end{abstract}

\maketitle


Quantum key distribution (QKD)  \cite{BB84,ekert1991} was proposed
as a method of achieving perfectly secure communications. Any
eavesdropping attempt by a third party will necessarily introduce an
abnormally high quantum bit error rate in a quantum transmission and
thus be caught by the users. With a perfect single photon source,
QKD provides proven unconditional security guaranteed by the
fundamental laws of quantum physics
\cite{securityproof,moresecurityproof}.

Most current experimental QKD set-ups are based on attenuated laser
pulses which occasionally give out multi-photons. Therefore, any
security proofs must take into account the possibility of subtle
eavesdropping attacks, including the PNS (photon number splitting)
attack  \cite{PNS}. A hallmark of those subtle attacks is that
they introduce a
photon-number dependent attenuation to the signal.

Fortunately, it is still possible to obtain unconditionally secure
QKD, even with (phase randomized) attenuated laser pulses, as
theoretically demonstrated by  \cite{ilm} and
Gottesman-Lo-L\"{u}tkenhaus-Preskill (GLLP)  \cite{GLLP}. However,
one must pay a steep price by placing severe limits on the distance
and the key generation rate. See also
 \cite{OtherPracticalSecurity}.

A key question is this: how can one extend the distance and key
generation rate of secure QKD? A brute force solution to this
problem would be to use a (nearly) perfect single photon source.
Despite much experimental effort  \cite{singlephoton}, reliable
perfect single photon sources are far from practical.

Another solution to increase the transmission distance and key
generation rate is to employ decoy states, using extra states of
different average photon number to detect photon-number dependent
attenuation. It has attracted great recent interests.
The decoy method was first discovered by
Hwang  \cite{hwang2003}. In  \cite{decoy}, we presented the first
rigorous security proof of decoy state QKD. We showed that the decoy
state method can be combined with standard GLLP result to achieve
dramatically higher key generation rates and distances. Moreover, we
proposed practical protocols with vacua or weak coherent states as
decoys. Subsequently, the security of practical protocols have been
analyzed by Wang  \cite{wang} and us  \cite{practical}.
See also \cite{harrington2005}. In
particular, we  \cite{practical} demonstrated theoretically the
clear practicality of decoy state QKD using only one decoy state. We
call such a protocol a one-decoy protocol.

However, until now, all decoy state QKD papers
 \cite{hwang2003,decoy,wang,practical,harrington2005} have been
theoretical and there has been no experimental demonstration. Here,
we present, for the first time, an experimental realization of decoy
state QKD.

We remark that additional errors will appear in experimental
implementation of a decoy state protocol. An example of a source of
additional errors is intensity modulation, which, as will be
discussed below, is required for the implementation of decoy state
QKD. Those additional errors will change the parameters and, thus,
the quantitative results in the simulations done in previous papers
 \cite{decoy,practical,wang}. Therefore, to quantify the advantage of
decoy state QKD in practice, it is crucial to perform a real
experiment and analyze the data obtained experimentally.


In our experiment, we use acousto-optic modulators (AOM) to achieve
polarization insensitive modulation, which is important for our
set-up. While already used in telecommunications, we believe that
this is the first time that AOMs have been introduced in a QKD
experiment. In summary, our experiment demonstrates a new
approach---decoy state QKD---with a new experimental
component---AOM---in QKD.

We will first discuss the GLLP result and how the decoy state method
can be combined with GLLP to achieve high key generation rate and
distance. The GLLP  \cite{GLLP} method can be used to prove the
security of QKD based on a phase randomized weak coherent state
source. With the GLLP method the secure key generation rate, which
is defined as the ratio of the length of the secure key to the total
number of signals sent by Alice, is given by  \cite{decoy}
\begin{equation}\label{R}
R\geq q\{-Q_\mu f(E_\mu)H_2(E_\mu)+Q_1[1-H_2(e_1)]\},
\end{equation}
where $q$ depends on the protocol \cite{q}, the subscript $\mu$ is
the average photon number per signal in signal states, $Q_\mu$  is
the gain  \cite{gain} of signal states, $E_\mu$ is the quantum bit
error rate (QBER) of signal states, $Q_1$ is the gain of the single
photon states (i.e., the probability that Alice generates exactly a
single photon which is finally detected by Bob) in signal states,
$e_1$ is the error rate of single photon states in signal states.
$f(x)$ is the bi-directional error correction rate
 \cite{brassard1994}, and $H_2(x)$ is the binary entropy function:
$H_2(x)=-x\log_2(x)-(1-x)\log_2(1-x)$. $Q_\mu$ and $E_\mu$ can both
be measured directly from experiments, while $Q_1$ and $e_1$ have to
be estimated (because we could not measure the photon number of each
pulse).

Owing to the loss in a fiber, its length determines the gains and
QBERs (as denoted by $Q_\mu, Q_1, E_\mu, e_1$) and therefore, the
key generation rate $R$. At the distance where the key generation
rate $R$ hits zero, the QKD protocol is no longer secure (with the
standard classical post-processing protocol which uses only one-way
classical communications).

Clearly, to estimate key generation rate, the main task is to
estimate a lower bound of $Q_1$ and an upper bound of $e_1$.
However, in non-decoy state approaches, the estimations are quite
poor. This is the reason why, with non-decoy state approaches, QKD
can be proven to be secure only at very limited key generation rate
and distance. While experimental QKD has been demonstrated at 122km
in telecom fibers  \cite{GYS}, most of the previous experiments do
not appear to satisfy the strict security analysis demanded in
non-decoy approaches  \cite{GLLP,insecurity}. Given that security is
the most crucial issue in QKD, this is a highly unsatisfactory
situation.

Fortunately, decoy state QKD comes to the rescue. As discussed
below, decoy state QKD allows dramatic improvement in our
estimations of $Q_1$ and $e_1$, compared to non-decoy approaches.
The basic idea of decoy state QKD is as follows: in addition to the
signal state with average photon number $\mu$, one introduces some
``decoy'' states with some other average photon numbers $\nu_i$ and
blends signal states with decoy states randomly on Alice's side
\cite{signals}. For instance, in a one-decoy state protocol
\cite{practical}, the average photon number of a decoy state is much
lower than that of the signal state. After Bob's acknowledgement of
receipt of signals, Alice broadcasts which pulses are signal states
and which are decoy states. Alice and Bob can, therefore, analyze
the statistical characteristics (i.e., transmittance and QBER) of
each type of signal separately. Since one assumes all
characteristics (except photon number distribution) of the signal
state and the decoy state are the same, Eve's eavesdropping attack
can depend on the actual photon number in each pulse, but she has no
knowledge which state (signal or decoy) the pulse is in. Eve's
attack will modify the characteristics (transmittance or QBER) of
decoy states and/or signal states and will be caught. For instance,
in a one-decoy state protocol, if Eve introduces a photon-number
dependent attenuation to the channel, then the transmittance of the
decoy state (which has a much lower average photon number than the
signal state) will generally be much lower than what Alice and Bob
would expect under normal operations. Note that decoy states are
used only for catching an eavesdropper, but not for key generation.
It has been shown \cite{decoy,wang,practical} that, in theory, decoy
state QKD can substantially enhance security and performance of QKD.

\begin{figure}[!t]\center
\resizebox{7.5cm}{!}{\includegraphics{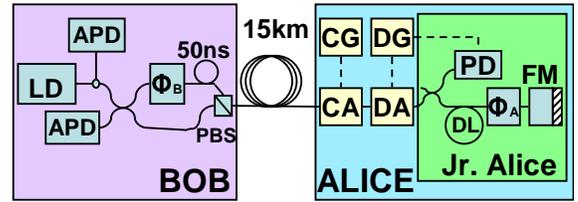}} \caption{Schematic
of the experimental set-up in our system. Inside Bob/Jr. Alice:
components in Bob/Alice's package of id Quantique QKD system. Our
modifications: CA: Compensating AOM; CG: Compensating Generator; DA:
Decoy AOM; DG: Decoy Generator. Original QKD system: LD: laser
diode; APD: avalanche photon diode; $\Phi_i$: phase modulator; PBS:
polarization beam splitter; PD: classical photo detector; DL: Delay
line; FM: faraday mirror. Solid line: SMF28 single mode optical
fiber; dashed line: electric cable.}\label{schematic}
\end{figure}

Before describing our experiment, we would like to point out that if
one already had a uni-directional QKD system  \cite{GYS,uniQKD} in
place, the implementation of decoy state QKD would have been much
easier: one can simply drive the laser source directly to various
power levels.

{\it Modified ``Plug and Play'' set-up:} Unfortunately, most
existing commercial QKD systems are bi-directional (``Plug \&
Play'') in the sense that Bob sends out a chain of strong signals to
Alice, who attenuates each signal to single photon level and
modulates (i.e., encodes quantum information on) it before sending
it back to Bob, who performs the measurement (i.e., decoding of
quantum information), after which a new chain of strong signals will
be sent to Alice. Therefore, the intensity modulation by Alice has
to be carefully synchronized with Bob's laser source.

Here, we show that, even with such a commercial
system (manufactured by id Quantique in our current set-up), one can
successfully implement decoy state QKD by making simple modifications.

{\it Our Experiment:} FIG. \ref{schematic} illustrates the schematic
of our system. The commercial QKD system consists of Bob and Jr.
Alice. In our experiment, Alice consists of Jr. Alice and four new
optical and electric components: an AOM (DA in FIG.~1), a function
generator (DG in FIG.~1), a compensating AOM (CA in FIG.~1) and a
compensating generator (CG in FIG.~1). Their functions will be
discussed below.

We implement the one decoy state protocol proposed by our group in
 \cite{practical} on top of the standard BB84 \cite{BB84} protocol. In
a one decoy state protocol, Alice must randomly modulate the
intensity of each signal to either signal state level or decoy state
level before sending it back to Bob. We add an AOM (DA in FIG.
\ref{schematic}) on Alice's side. A function generator (DG in FIG.
\ref{schematic}) controls the transmittance of the AOM.

A critical issue in our set-up is to compensate the phase shift due
to our AOM. Since the Mach-Zehnder Interferometer (MZI) on Bob's
side is asymmetric, the frequency shift introduced by AOM causes a
significant phase shift between the two pulses that go through
different arms of MZI. To compensate this phase shift, another AOM,
the ``Compensating AOM'' (CA in FIG. \ref{schematic}) is employed to
make the total phase shift multiples of $2\pi$. This AOM is driven
by the second function generator, ``Compensating Generator'' (CG in
FIG. \ref{schematic})  \cite{CACG}.

{\it Optimization of Parameters:} We perform a numerical
simulation \cite{practical} with parameters of our set-up
\cite{ExpPara} and optimally set $\mu$ and $\nu$ to 0.80 and 0.12
photons respectively. The actual distribution of the states is
produced by an id Quantique Quantum Random Number Generator. Around
10\% of the signals are assigned as decoy states, which is optimal
according to numerical simulation \cite{practical}. This random
pattern is generated and loaded to the Decoy Generator before the
experiment \cite{decoyprofile}.

Except for the modulation period, the transmittance of Decoy AOM (DA) is
set to maximum. As the classical detector (PD in FIG. \ref{schematic}) inside Alice
detects the first pulse from Bob, it triggers the Decoy Generator (DG).
The Decoy Generator (DG) will then hold a delay time $t_d$ before
outputting the random pattern to modulate different states. The
Compensating AOM (CA) is used only for the purpose of phase compensation.
Thus, its transmittance is set to be constant.

Recall that each signal in a ``Plug and Play'' set up consists of
two time-separated pulses. To keep visibility high, the two pulses
of the same signal must be attenuated equally, which means the delay
time must be very precise. In our experiment, the delay time $t_d$
was determined with an accuracy of 10ns.

In our experiment, a total of $N=105$M raw bits (including both
signal states and decoy states) were sent using quantum key
distribution from Alice to Bob. The transmitting time was less than
four minutes.

After the transmission of all the $N$ signals, Bob announced which
signals had actually been received by him and in which basis. Alice
broadcasted to Bob the distribution of decoy states as well as basis
information. We assume Alice and Bob announced the measurement
outcomes of all decoy states as well as a subset of the signal
states. From those experimental data, Alice and Bob then determined
$Q_\mu$, $Q_\nu$ $E_\mu$, and $E_\nu$, whose values are now listed
in Table~\ref{expresult}.

\begin{table}[!b]\center
\begin{tabular}{c c |c c| c c}
\hline

Para. & Value & Para. & Value & Para. & Value\\

\hline $Q_{\mu}$ &$8.757\times10^{-3}$ &$E_{\mu}$
&$9.536\times10^{-3}$ &$q$&0.4478\\
 $Q_{\nu}$ &$1.360\times10^{-3}$ &$E_\nu$&$2.689\times10^{-2}$
&$f(E_\mu)$ \cite{brassard1994}& $\leq$1.22\\
\hline

\end{tabular}
\caption{Experimental data and some parameters we used in our
experiment. As required by GLLP \cite{GLLP}, bit values for double
detections are assigned randomly by the quantum random number
generator.}\label{expresult}
\end{table}

{\it Analysis of Experimental Results:} Alice and Bob have to derive
a lower bound on the key generation rate, $R$, by applying the
theory of one decoy state protocol to their experimental data. To
begin, we discuss the theory of one decoy state protocol. The
one-decoy state protocol was first proposed in  \cite{decoy} and
analyzed in  \cite{practical}. In such a protocol, only one decoy
state is used (in principle, more decoy states might increase key
generation rate) whose average photon number is $\nu$. The
transmittance/gain of the decoy state $Q_\nu$ and its error rate
$E_\nu$ could also be acquired directly from experiments. Taking
statistical fluctuations into account, the lower bound of $Q_1$, and
the upper bound of $e_1$ are given by \cite{practical}

\begin{equation}
\begin{aligned}
Q_1 &\ge Q_1^L = \frac{\mu^2e^{-\mu}}{\mu\nu-\nu^2}(Q_\nu^L
e^{\nu}-Q_\mu e^\mu\frac{\nu^2}{\mu^2}-E_\mu Q_\mu e^\mu
\frac{\mu^2-\nu^2}{e_0\mu^2})\\
e_1 &\le e_1^U = \frac{E_\mu Q_\mu}{Q_1^{L}},
\end{aligned}
\end{equation}
in which
\begin{equation}
Q_\nu^L = Q_\nu(1-\frac{u_\alpha}{\sqrt{N_\nu Q_\nu}}),
\end{equation}
where $N_\nu$ is the number of pulses used as decoy states
 \cite{fluctuation}, and $e_0$ (=1/2) is the error rate for the
vacuum signal and therefore the lower bound of key generation rate
is
\begin{equation}
R \ge R^L=q\{-Q_\mu f(E_\mu)H_2(E_\mu)+Q_1^L[1-H_2(e_1^U)]\}
\end{equation}

In our analysis of experimental data, we estimated $e_1$ and $Q_1$
very conservatively as within 10 standard deviations (i.e.,
$u_\alpha$=10), which promises a confidence interval for statistical
fluctuations of $1-1.5\times10^{-23}$.

The experimental results listed in Table \ref{expresult} are the
inputs for Eqs. (2-4), whose output is a lower bound of the key
generation rate, as shown in Table \ref{yqerbound}. Even with our
very conservative estimation of $e_1$ and $Q_1$, we got a lower
bound for the key generation rate $R^L=3.6\times10^{-4}$ per pulse,
or 165bits/s, which means a final key length of about
$L=NR\simeq38$kbit. We also calculated $R_{perfect}$, the
theoretical limit from the case of infinite data size and infinite
decoy states protocol, by using Eq. (1). We remark that our lower
bound $R^L$ is indeed good because it is roughly $1/4$ of
$R_{perfect}$. This fact suggests that it is not necessary, or
rather, not economical to use either a very large quantity of data
or a lot of different decoy states.

\begin{table}[!b]\center

\begin{tabular}{c c|c c}
\hline Para. & Value & Para. & Value\\
\hline $Q_1^L$  & $2.140\times10^{-3}$ & $R^L$    & $3.588\times10^{-4}$\\
$e_1^U$  & $3.902\times10^{-2}$ &$R_{perfect}$ &
$1.418\times10^{-3}$\\
 \hline

\end{tabular}

\caption {The lower bounds of $Q_1$, $R$ and the upper bound of
$e_1$. The values are calculated from Eqs. (2-4). As a comparison,
we also gave the theoretical limit, $R_{perfect}$. It represents the
situation of infinitely long data size and infinitely many decoy
states. Our result shows that even a simple one decoy state protocol
can achieve one fourth of the theoretical limit.}\label{yqerbound}
\end{table}

Based on the method described in  \cite{GLLP,decoy,practical}, we
carefully performed numerical simulations with \cite{ExpPara}. We
found that without decoy method, no matter what value of $\mu$ we
choose or how large the data size we use, the key generation rate,
$R$, will hit zero at only 9.6km. In other words, at 15km, not even
a single bit could be shared between Alice and Bob with guaranteed
security. In contrast, our numerical simulations show that, with
decoy states, our QKD set-up can be made secure over 50km, which is
substantially larger than the secure distance (9.6km) without decoy
states.

In summary, we have performed the first experimental demonstration
of decoy state QKD, over 15km of Telecom fibers. Our experiment
shows that, with rather simple modifications (by adding commercial
acousto-optic modulators, AOM) to a commercial QKD system, decoy
method allows us to achieve much better performance with
substantially higher key generation rate and longer distance than is
otherwise possible. We conclude that, with careful conceptual design
and optimization, decoy state QKD is easy to implement in
experiments. It is, therefore, ready for immediate commercial
applications.

We thank enlightening discussions with colleagues including C. H.
Bennett, D. Bethune, G. Brassard, F. Dupuis, J. Harrington, H. J.
Kimble, L. LeBlanc, D. W. C. Leung, N. L\"{u}tkenhaus, L. McKinney,
J. Preskill, C. Rose, K. Tamaki, X.-B. Wang and Z. Yuan. We
particularly thank G. Ribordy for his generous help in our
experiment. Financial support from Connaught, NSERC, CRC Program,
CFI, id Quantique, OIT, PREA, CIPI, and the University of Toronto is
gratefully acknowledged. H.-K. Lo also thanks travel support from
the IQI at Caltech through the NSF under grant EIA-0086038.


\begin{thebibliography}{00}


\bibitem{BB84}
C. H. Bennett,  G.Brassard, Proceedings of \emph{IEEE International
Conference on Computers, Systems, and Signal Processing}, (IEEE,
1984), pp. 175-179.
\bibitem{ekert1991}
A. K. Ekert, \emph{Phys. Rev. Lett.} \textbf{67} 661 (1991)

\bibitem{securityproof}
D. Mayers, \emph{J. of ACM} \textbf{48}, 351 (2001); H.-K. Lo, H. F.
Chau, \emph{Science}, \textbf{283}, 2050 (1999); E. Biham \emph{et
al.} Proceedings of \emph{the Thirty-Second Annual ACM Symposium on
Theory of Computing (STOC'00)} (ACM Press, New York, 2000), pp.
715-724; P. W. Shor, J. Preskill, \emph{Phys. Rev. Lett.}
\textbf{85}, 441, (2000)
\bibitem{moresecurityproof}
K. Ekert, and B. Huttner, {\it J. of Modern Optics} {\bf 41}, 2455
(1994); D. Deutsch {\it et al.}, {\it Phys. Rev. Lett.} {\bf 77},
2818 (1996); Erratum: {\it Phys. Rev. Lett.} {\bf 80}, 2022 (1998).



\bibitem{PNS}
Eve can selectively surpress all the single photon signals from
Alice, and split all the multi photon signals, keeping one copy
herself and send the other copy to Bob. In this way, Eve could have
an identical copy of what Bob processes, thus breaking the security
of BB84 protocol. Although such attacks may appear to be beyond
current technology, the first rule in cryptography is: never
underestimate the determination and ingenuity of your opponents in
breaking your codes.

\bibitem{ilm}
H. Inamori, N. L\"{u}tkenhaus, D. Mayers, in press (available at
http://arxiv.org/abs/quant-ph/0107017)

\bibitem{GLLP}
D. Gottesman \emph{et al.} \emph{Quantum Information and
Computation} \textbf{4}, 325 (2004)

\bibitem{OtherPracticalSecurity}
M. Koashi, available at http://arxiv.org/abs/quant-ph/0403131 ;
V. Sacarani \emph{et al.} \emph{Phys. Rev. Lett.} \textbf{92}
057901, K. Inoue \emph{et al.} \emph{Phys. Rev. A} \textbf{71}
042305, M. Curty \emph{et al.} \emph{Phys. Rev. A} \textbf{69}
042321
\bibitem{singlephoton}
J. McKeever, \emph{et al}. \emph{Science} \textbf{303}, 1992 (2004);
Z. L. Yuan, \emph{et al}, \emph{Science} \textbf{295}, 102 (2002);
M. Keller \emph{et al.} \emph{Nature} \textbf{431}, 1075 (2004)


\bibitem{hwang2003}
W.-Y. Hwang, \emph{Phys. Rev. Lett}. \textbf{91}, 057901 (2003)
\bibitem{decoy}
H.-K. Lo, in {\it Proceedings of IEEE ISIT 2004}, p. 137;
H.-K. Lo, X. Ma,  K. Chen, \emph{Phys. Rev. Lett.} \textbf{94}
230504 (2005).

\bibitem{wang}
X. -B. Wang, \emph{Phys. Rev. Lett.} \textbf{94} 230503 (2005); X.
-B. Wang, \emph{Phys. Rev. A} \textbf{72} 012322 (2005)

\bibitem{practical}
X. Ma \emph{et al.} \emph{Phys. Rev. A} \textbf{72} 012326 (2005)

\bibitem{harrington2005}
J. W. Harrington \emph{et al.} in press (available at
http://arxiv.org/abs/quant-ph/0503002)

\bibitem{q}
Our experimental implementation is based on BB84 protocol. In the
cases that Alice and Bob use the same basis, $N_\mu^S$  pulses are
used as signal states. Factor $q$ in Eq. (\ref{R}) is thus given by
$q=N_\mu^S/N$.

\bibitem{gain} The gain is defined to
be the ratio of the number of receiver Bob's detection events to the
number of signals emitted by sender Alice in the cases where Alice
and Bob use the same basis.
\bibitem{brassard1994}
G. Brassard, L. Salvail,  \emph{Advances in Cryptology EUROCRYPT
'93}, Vol. 765 of \emph{Lecture Notes in Computer Science},
(Springer, Berlin, 1994), pp. 410-423.

\bibitem{GYS}
C. Gobby, Z. L. Yuan, A. J. Shields, \emph{Appl. Phys. Lett.}
\textbf{84} 3762 (2004)
\bibitem{insecurity}
Many existing QKD experiments take an ad hoc value of 0.1 for the
average photon number of the signal. Besides, few of them have
considered the most general eavesdropping attack allowed by quantum
mechanics.

\bibitem{signals} Here, we assume that the source emits coherent
states and the phases of all states
(signals or decoys) are randomized. Therefore, the photon
number of the signal state follows a Poissonian distribution
with a parameter $\mu$ whereas
that of a decoy state follows a Poissonian distribution
with a parameter $\nu_i$.

\bibitem{uniQKD}
The same strategy fails miserably for a bi-directional QKD system
because Eve can easily monitor the intensity of the ancillary strong
signal originated from Bob.

\bibitem{CACG}
Given a variable frequency driver for DA (which is also commercially
viable), CA and CG could be taken off.


\bibitem{ExpPara}
Laser $\lambda=1550$nm at 5MHz, background rate
$Y_0=2.11\times10^{-5}$, fiber loss 0.21dB/km, Bob's quantum
efficiency $\eta_{Bob}$=0.0227, detector error rate
$e_{detector}=8.269\times10^{-3}$.

\bibitem{decoyprofile}
In principle, for perfect security, a new random pattern should be
chosen for each frame. For ease of implementation, in our
experiment, the same pattern was reused for every frame. Note,
however, that each signal within a frame was still modulated
individually.

\bibitem{fluctuation}
This is because in our experiment, the yield for the cases that
Alice and Bob use the same basis is the same as that for the cases
that they use different basis.


\end{thebibliography}
\end{document}